\documentclass[sn-nature]{sn-jnl}

\usepackage{graphicx}%
\usepackage{multirow}%
\usepackage{amsmath,amssymb,amsfonts}%
\usepackage{amsthm}%
\usepackage{mathrsfs}%
\usepackage[title]{appendix}%
\usepackage[dvipsnames]{xcolor}%
\usepackage{textcomp}%
\usepackage{manyfoot}%
\usepackage{booktabs}%
\usepackage{algorithm}%
\usepackage{algorithmicx}%
\usepackage{algpseudocode}%
\usepackage{listings}%
\usepackage{tikz}
\usepackage{pgfplots}
\usepackage{xurl}

\theoremstyle{thmstyleone}%
\theoremstyle{thmstyletwo}%

\theoremstyle{thmstylethree}%

\begin{document}

\title[On the Fast Track to Full Gold Open Access]{On the Fast Track to Full Gold Open Access}


\author[]{\fnm{Robert} \sur{Kudeli\'c}}\email{robert.kudelic@foi.unizg.hr}

%

\affil[]{
	\orgname{University of Zagreb Faculty of Organization and Informatics}, \orgaddress{
		\country{Republic of Croatia}}}

%


\abstract{The world of scientific publishing is changing; the days of an old type of subscription-based earnings for publishers seem over, and we are entering a new era. It seems as if an ever-increasing number of journals from disparate publishers are going Gold, Open Access that is, yet have we rigorously ascertained the issue in its entirety, or are we touting the strengths and forgetting about constructive criticism and careful weighing of evidence? We will therefore present the current state of the art, in a compact review/bibliometrics style, of this more relevant than ever hot topic, including challenges and potential solutions that are most likely to be acceptable to all parties. Suggested solutions, as per the performed analysis, at least for the time being, represent an inclusive publishing environment where multiple publishing models are competing for a piece of the pie and thus inhibiting each other's flaws. The performed analysis also shows that there seems to be a link between trends in scientific publishing and tumultuous world events, which in turn has a special significance for the publishing environment in the current world stage---implying that academy publishing has potentially now found itself at a tipping point of change.}




\maketitle

\section{Introduction}\label{sec1}

Once, not even that long ago, subscription-based scientific publishing was the norm. \cite{Bjoerk2012} The authors would write a paper after the hard work of research and, with assurance, send the paper to the selected journal for review. Some time would pass, and most likely than not, corrections to the paper would be requested and made, as the wisdom is in the advice of many. After this long-established iterative process would come to a near conclusion, a decision would be made, the editor would have spoken, let us say accepted for publication, the copy editing process would then proceed, the paper would be published, and the publisher, together with the authors, would be happy. All was well.

Subscriptions were paid by organizations that needed knowledge from the published articles, typically universities and governments. This kind of model found in scientific publishing \cite{Bjoerk2012} is a standard in almost every branch of human existence; there is a producer of a product, which invests in the product, and there is a customer, which has made no investment in the product but is in need of it and therefore naturally pays the deserving price to the producer \cite{Kirzner1978,Ebner2010}. With a catch in scientific publishing, where the author most likely received no compensation from the publisher (and typically neither did editors nor reviewers) \cite{Andrei2020}, in essence, publishers typically held onto all the proceeds, a strange twist but nonetheless a practice.

The problem was, however, pinpointed to this model of conducting scientific publishing, namely that there are those that truly can't afford the subscription as they are expensive, with the freedom of science also being a chiming sound, and so a movement was started, in a more prominent form this time, which touted Open Access\footnote{Enabled by innovation in digital technology. \cite{Ginsparg2011}} publishing. \cite{White1976,Laakso2011,OGara2019} Through time, various forms of this kind of publishing were designed and put into practice, with the basic idea that the final version of the article should be free for the reader; this kind is typically called Gold Open Access (although there are other kinds) \cite{Laakso2011} and is the model most aspired to; thus, publishers business model over the years and decades changed and is still changing. An increasing number of articles are continually being free to the reader, with Article Processing Charges coming into place \cite{Burchardt2014}, so as to fully or in part replace the tried and true subscription model, which indeed was not beyond improvement yet has served science well and for a long time.

Therefore, to further research and discussion \cite{Zhang2022}, to review the most important information, to present new information and insight, and to suggest a model of scientific publishing that, according to current findings, best ensures open science from all angles, we present the current research.

\section{Relevant Literature}\label{sec2}

Many supported an Open Access model \cite{Zhao2014,Bjoerk2017}, with publishers trying to bring even more journals to such a model every year, and especially to the Gold Open Access typically hybridized by the option of the subscription-based model, with the transition to Full Gold Open Access paving its way (a type of Open Access model where the only option to publish is by paying a publication fee, that is Article Processing Charge). \cite{Bjoerk2012,Sivertsen2019}

From a business standpoint, the financial construction is sound; there is a cut in funds on one side, but that will be covered by the authors or funding institutions on the other side, and thus the publishers will survive, an important factor without a doubt, as the publishers are an integral part of academia and the scientific community. Yet the dreadfully expensive subscriptions were replaced by typically expensive, or at least not that inexpensive, Article Processing Charges. With the natural question in the air: What is the total sum in terms of pluses and minuses, and what are the constraints? \cite{Beall2013}

With the Article Processing Charge, by having in mind Gold Open Access if that is the future, the author is no longer so assured after the research is over and the paper written, as even if the research is of excellent quality, the great wall of Article Processing Charge is casting a shadow, and it might be quite a long one. \cite{Burchardt2014} Article Processing Charges are as opposed to typical conference fees (common entry paywall of the academia) steep; they range from $ 15 $ dollars, all the way to $ 10183 $ dollars, with a mean of $ 2987 $ and a standard deviation of $ 1352 $ dollars \cite{Budzinski2020}, which can be a challenge even for scientists coming from a first-world country \cite{Du2022,Andrei2020}---and if authors can't pay the charge, if the charge is not waived in some way, the publication will not happen, and the science will be crushed.

The height of Article Processing Charge is clearly a concern \cite{Zhang2022,Shu2023}; it seems that the Open Access model is ca. equivalent in price or not that much more inexpensive, quite a lackluster performance considering the hype, and if the situation is like this now, what will happen when there will be no competition to press the brakes? What if the only model will be the Full Gold Open Access (it seems that publishers are indeed planning for this \cite{ACM2023,Elsevier2023,Springer2021,Packer2023})? \cite{Andrei2020,Du2022}

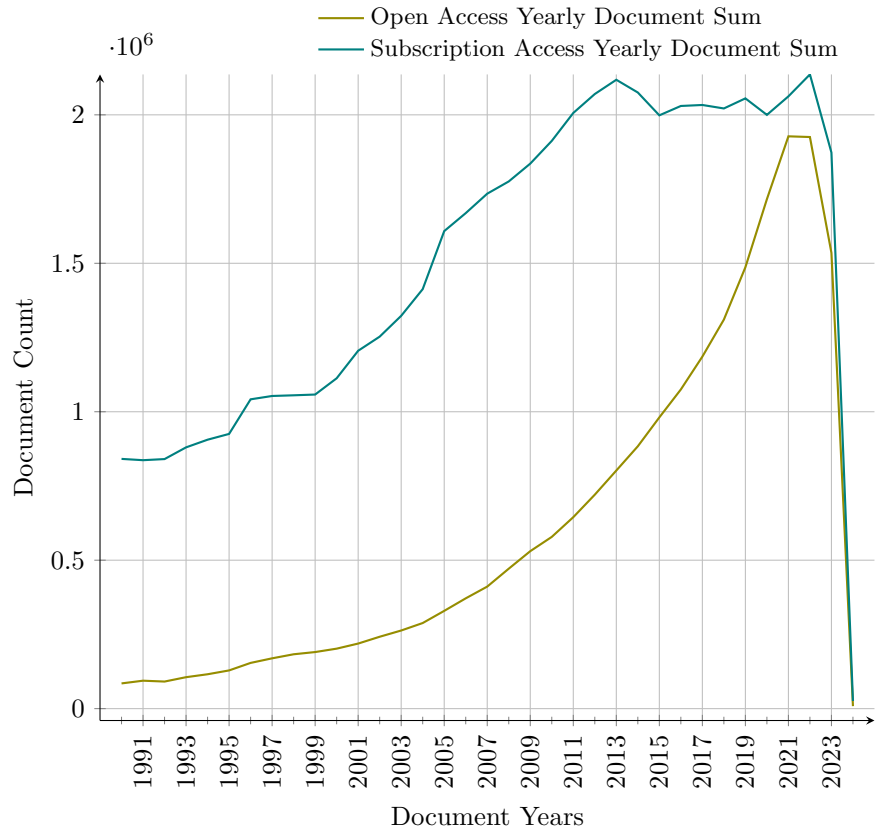
\begin{figure}[h]%
	\centering
	\begin{tikzpicture}[]
		\begin{axis}[
			ylabel near ticks,
			xlabel near ticks,
			minor xtick={1990, 1992, ..., 2024},
			xtick={1991, 1993, ..., 2023},
			xmin=1989, xmax=2025,
			ymin=-40000, 
			xlabel={Document Years},
			ylabel={Document Count},
			grid=major,
			x tick label style={/pgf/number format/1000 sep=},
			width=.9\textwidth,
			xticklabel style={rotate=90},
			axis lines=left,
			legend style={at={(.62,1.12)},
				anchor=north,legend columns=1,legend cell align={left},draw=none},
			]
			\addlegendentry{\scriptsize Open Access Yearly Document Sum}
			\addplot[thick, color=olive]
			table[x=Godina, y=Broj radova, col sep=comma]{OA_Radovi_po_tipu.nlo};\label{dokOA}
			\addlegendentry{\scriptsize Subscription Access Yearly Document Sum}
			\addplot[thick, color=teal]
			table[x=Godina, y=Broj radova, col sep=comma]{SB_Radovi_po_tipu.nlo};\label{dokSB}
		\end{axis}
	\end{tikzpicture}
	\caption{Open Access (OA) and Subscription Access (SA) Documents by Publication Year: The last two years obviously do not have a complete dataset. Data was acquired from Scopus, which is the "world's largest abstract and citation database of peer-reviewed research literature" \cite{Elsevier2023a}. No filter was set on document type, and as such, this is a complete and up-to-date outlook. Up until recently, SA publishing has commanded a substantial lead, yet the steady and especially recent explosion of OA documents has placed both publishing models close to one another. Such a steep growth in OA publishing has, from ca. 2013, produced, it seems, a standstill in the SA model; however, from ca. 2021, the OA model has perhaps reached its peak as well, although it is still too soon to tell. Notwithstanding such an instance, it seems that we are at a tipping point, at a crossroads from which we will see a different turn of events. For a historical and more in-depth overview, one should consult section \ref{sec7} of the article. Documents for the year 2024 are the ones with a planned date.}\label{fig:OaSb}
\end{figure}

This kind of Gold Open Access model, and those alike to it, might be defended on the ground that the authors do not have to, or are not, paying the charge themselves, with a library or some other funder covering the expense. But are we suggesting that every university, college, faculty, small and medium-sized enterprise, etc. has the funds to cover the Article Processing Charge without which one can't publish a scientific discovery? \cite{Andrei2020} To discern the issue more clearly, we can look at the global survey on Open Access books\footnote{Inclusion of such an analysis makes the present article more complete, gives an insight into a branch of publishing similar in some instances to articles publishing (e.g. scientific monographs), and presents another publishing area with which one can compare and reason about.} from academic book authors, which found that $ 81\% $ of Open Access book authors and $ 55\% $ of non-Open Access book authors agree or strongly agree "that all future scholarly books should be" Open Access. \cite{Pyne2019} Thus the number of authors taking a different stance is not insignificant; that is, $ 19\% $ of authors from Open Access book authors and $ 45\% $ of authors from non-Open Access book authors are in the other part of the spectrum. Speaking in absolute terms, out of the 2542 respondents, ca. $ 80\% $ (2037) are non-Open Access book authors, therefore making a case argued by non-Open Access book authors far stronger than that of Open Access book authors. \cite{Pyne2019} This research has also found that typical barriers to choosing Open Access, given by non-Open Access book authors, are inability to find funds (representing a serious hurdle, as a high proportion of $ 47\% $ of both types of authors stated that they "didn't have funding for their last book"), low quality perception, low awareness, and lack of willingness to pay, while the top three motivations for choosing Open Access, stated by Open Access book authors, are beliefs in: larger readership, openness of research, and higher citation count. \cite{Pyne2019} For a comparison of the Open Access and Subscription Access document output trends, one can consult Figure~\ref{fig:OaSb}.

Yet there are other issues following: what about individuals who are conducting research, perhaps in their private time, and without compensation, will they be paying the Article Processing Charge so as to publish their paper? \cite{Beall2013} Evidently, the situation is not as clear-cut as one might perceive at a glance. If there is no way to publish in a respectable venue without paying for publishing, then we are denying quality research for the reason of price. This does not sound like belonging to science, and it isn't, as Article Processing Charge should not determine whether something can be published or not, yet that long shadow is saying, either pay or leave, the criteria has changed from meritocracy to paytocracy. Will science close the door to science? Is the future of science only that which is paid for? Is this really independent research and traveling to any place to which the evidence leads?

As unpleasant as the critique may seem, scientific work and science in general depend on a number of very specific postulates (freedom of research, unblocked access to publishing venues, independent and fair peer review and editorial process, certainty that accepted research will be published speedily). With this disturbance, science is in serious danger. The fact is that science needs to be protected; the status quo in science is neither an acceptable result nor a conclusion. If SA publishing became unbearable for the reason of subscription price, OA publishing is becoming unbearable for the reason of APC price \cite{Zhang2022}, and moving from one extreme to another will not be constructive.

One of the places where paytocracy is horrifyingly evident are predatory journals \cite{Grudniewicz2019} and publishers which have every incentive to publish as high number of articles as they can, since the reader is not paying anymore and has became irrelevant, a general flaw of the Open Access publishing\footnote{Typical models are (Full) Gold Open Access, Hybrid Open Access, Transformative Open Access, Green Open Access, Diamond/Platinum Open Access. \cite{Laakso2011,Packer2023}} \cite{Beall2012}---with most notable characteristics of such endeavors probably being large number of special issues, contacting large number of authors and soliciting papers (potentially with discounted or heavily discounted Article Processing Charge), unusually short reviewing process, unusually high number of published articles, using scientists as a facade in occupied positions, unusually high repeated authorship, excessive editor authorship, low quality research, unofficial impact factor, contacting authors of a work published as a preprint excessive number of times so as to publish it for a certain publisher or in a certain publication, etc. \cite{Butler2013,Gilbert2009,Beall2012,Beall2013,Jalalian2013,Elliott2012,Beall2015,Eriksson2016,Shen2015}

Yet it is so difficult to prove\footnote{More on predatory element one can also find on Research Square. \cite{Prater2023}}, and so difficult to show that there is a predatory element in a journal or with a specific publisher, the veil is sometimes so perfected that it takes a substantial effort from multiple experts to reach a conclusion that yes, there is something very strange going on here, we need to publish elsewhere. \cite{Beall2012}

Everything has a price, and zero-cost publishing is unrealistic \cite{Hutson2021}, yet the current price is high---too high, many would argue, but why are publishers not budging? Why are these very expensive models thriving? Because there is a market for them, and as long as that is the case, little to nothing will change---parties involved have to stop paying these towering prices, and funders need to stop rolling out checks to cover these huge expenses, as talk alone will not go far. Self-balancing and competitive models need to coexist so as to accommodate all facets of science.

\section{Methodology}\label{sec6}

This research and the article represent a focused review and bibliometrics on the current state of, although not exclusively, scientific journal publishing and, more specifically, on the accompanying publishing models. In order to achieve the research goal, both Google Scholar and the World Wide Web were searched for the most focused and results-oriented articles dealing with scientific publishing and providing solutions to its problems.

The priority was given to high-impact sources, sources that are tackling issues in a direct manner, sources that are of a more recent date, and sources that are bringing significant new information and insight. This article, however, does not represent an exhaustive survey of the problem, since the goal is not to bring to light everything said about the issue but to narrow in on that literature that will constructively and in a state-of-the-art manner expound and expand knowledge and progress. Thus, articles repeating already established information are not necessarily included in the analysis, as this would overflow the article and diverge from the flow and the tackled issues; while the search was conducted for English-language articles.

The article, therefore, represents the most recent and state-of-the-art report on, primarily, scientific journal publishing models, their challenges, consequences, historical indicators, and immediate solutions to issues raised by the surveyed corpus of knowledge and our own analysis of the issues.

Furthermore, the data presented in the figures of the article is obtained from the Scopus database, and this has been conducted in the following manner. First, one needs to access the Scopus advanced document search, where the following query string needs to be searched for: "PUBYEAR AFT 1500"; by this act, Scopus will produce, as a result, its entire document corpus through the entire lifespan of the Scopus database. Now, one can filter through documents by various open access categories (all open access, green, gold, bronze, and hybrid gold). Then, by selecting "Individual" in the "Year" filter and then selecting "Show all," one can obtain the data for all the years that Scopus has data on. By excluding "All open access" documents, one can also obtain documents that are currently categorized as not published via the open access model.

Figures in the article that present data for OA are using the data from the category "All open access" under the filter "Open access." Such aggregation allows us to combine the available data, emphasize trends that would potentially be less visible and more difficult to detect, project additional data into the past, and thus indicate the potential historical turn of events. Nevertheless, there are also constraints on the analysis and the data. Namely, the farther down the timeline one makes such a projection, the more likely it is that an approximation will be farther from the possible real state at that particular time. While if one considers the data, category "Green" consists of articles that were made available in various ways and were likely predominantly, perhaps exclusively, published via the subscription model and then later on made available, and category "Gold" holds data values all the way back to the year 1874.

Yet, when we have analyzed all those categories separately, the trends that have been revealed were similar to almost identical to the trends that the combined data in the figures of the article reveals---indicating that the assumption of the analysis is correct and that the presented data represents a solid foundation for conducting analysis and producing conclusions and implications under the goal of the research.

\section{Historical Outlook}\label{sec7}

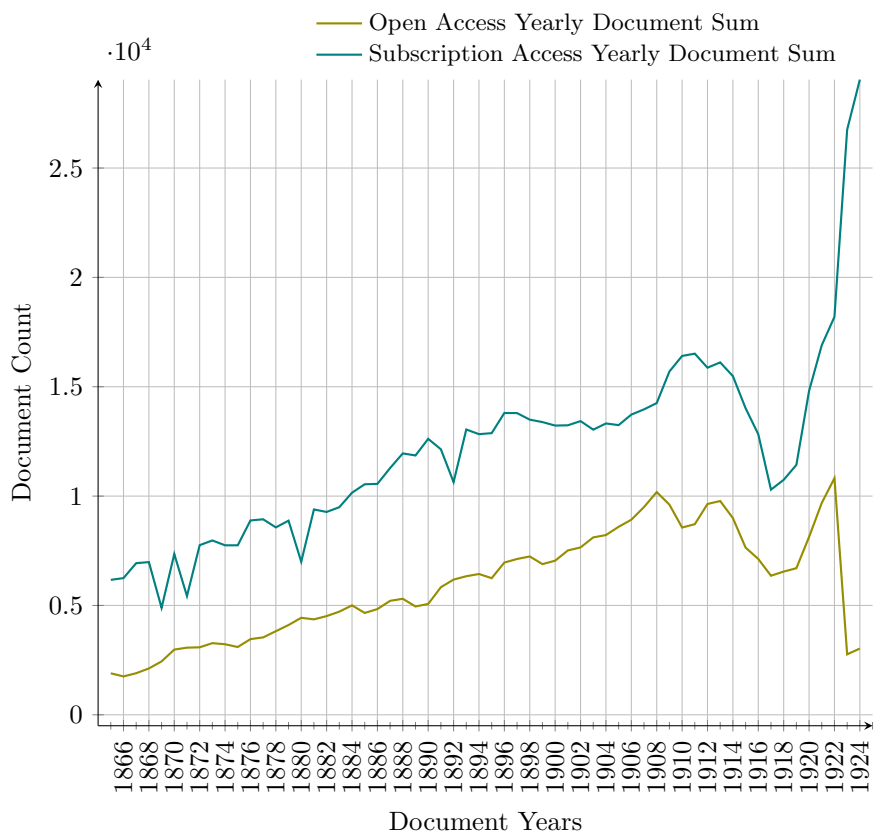
\begin{figure}[h]
	\centering
	\begin{tikzpicture}[]
		\begin{axis}[
			ylabel near ticks,
			xlabel near ticks,
			minor xtick={1865, 1867, ..., 1925},
			xtick={1866, 1868, ..., 1924},
			xmin=1864, xmax=1925,
			ymin=-500, 
			xlabel={Document Years},
			ylabel={Document Count},
			grid=major,
			x tick label style={/pgf/number format/1000 sep=},
			width=.9\textwidth,
			xticklabel style={rotate=90},
			axis lines=left,
			legend style={at={(.62,1.12)},
				anchor=north,legend columns=1,legend cell align={left},draw=none},
			]
			\addlegendentry{\scriptsize Open Access Yearly Document Sum}
			\addplot[thick, color=olive]
			table[x=Godina, y=Broj radova, col sep=comma]{1_OA_Radovi_po_tipu.nlo};\label{1dokOAsupp1}
			\addlegendentry{\scriptsize Subscription Access Yearly Document Sum}
			\addplot[thick, color=teal]
			table[x=Godina, y=Broj radova, col sep=comma]{1_SB_Radovi_po_tipu.nlo};\label{1dokSBsupp1}
		\end{axis}
	\end{tikzpicture}
	\caption{Open Access (OA) and Subscription Access (SA) Documents by Publication Year---Part One, from 1865 to 1924. Data was acquired from Scopus, which is the "world's largest abstract and citation database of peer-reviewed research literature" \cite{Elsevier2023a}. No filter was set on document type, and as such, this is a complete and up-to-date outlook.}\label{fig:1OaSbSupp1}
\end{figure}

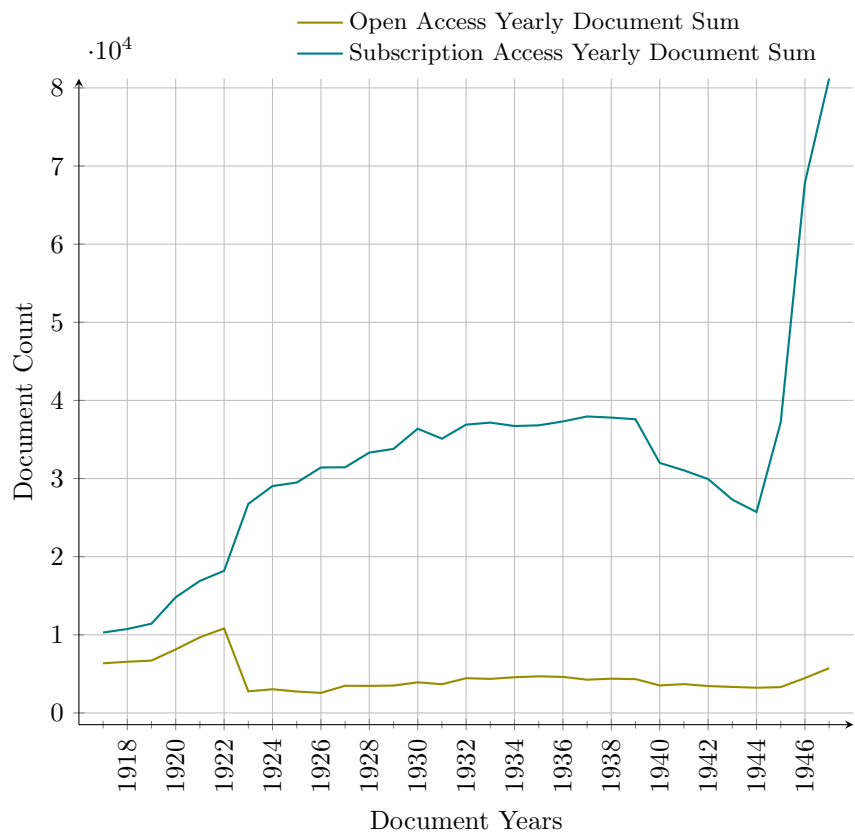
\begin{figure}[b]
	\centering
	\begin{tikzpicture}[]
		\begin{axis}[
			ylabel near ticks,
			xlabel near ticks,
			minor xtick={1917, 1919, ..., 1948},
			xtick={1918, 1920, ..., 1947},
			xmin=1916, xmax=1948,
			ymin=-1500, 
			xlabel={Document Years},
			ylabel={Document Count},
			grid=major,
			x tick label style={/pgf/number format/1000 sep=},
			width=.9\textwidth,
			xticklabel style={rotate=90},
			axis lines=left,
			legend style={at={(.62,1.12)},
				anchor=north,legend columns=1,legend cell align={left},draw=none},
			]
			\addlegendentry{\scriptsize Open Access Yearly Document Sum}
			\addplot[thick, color=olive]
			table[x=Godina, y=Broj radova, col sep=comma]{2_OA_Radovi_po_tipu.nlo};\label{1dokOAsupp2}
			\addlegendentry{\scriptsize Subscription Access Yearly Document Sum}
			\addplot[thick, color=teal]
			table[x=Godina, y=Broj radova, col sep=comma]{2_SB_Radovi_po_tipu.nlo};\label{1dokSBsupp2}
		\end{axis}
	\end{tikzpicture}
	\caption{Open Access (OA) and Subscription Access (SA) Documents by Publication Year---Part Two, from 1917 to 1947. Data was acquired from Scopus, which is the "world's largest abstract and citation database of peer-reviewed research literature" \cite{Elsevier2023a}. No filter was set on document type, and as such, this is a complete and up-to-date outlook.}\label{fig:1OaSbSupp2}
\end{figure}

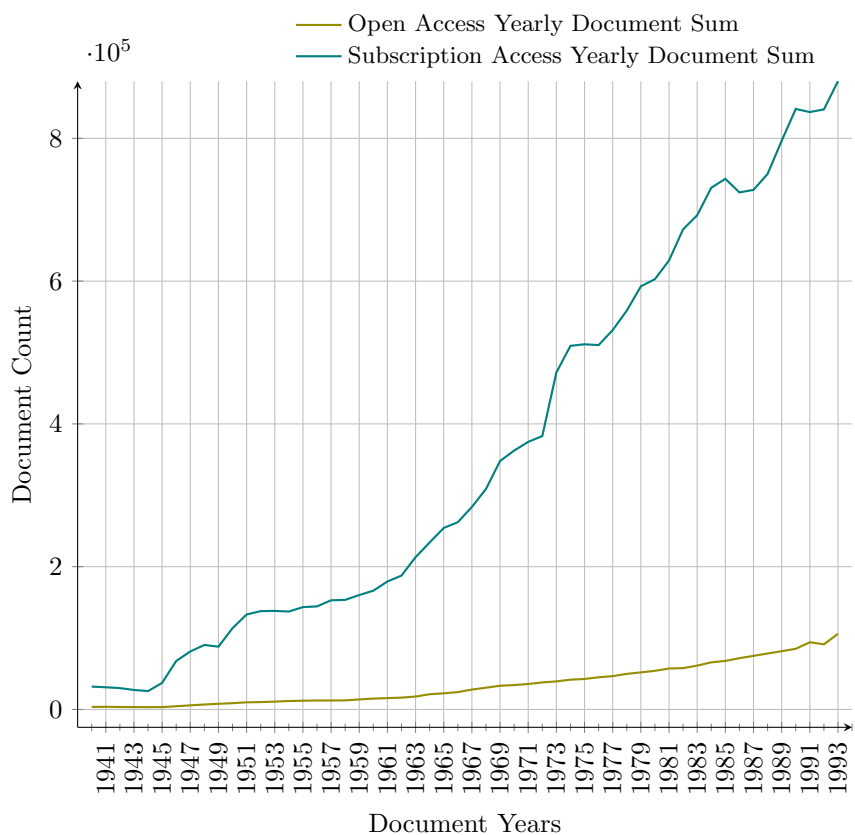
\begin{figure}[b]
	\centering
	\begin{tikzpicture}[]
		\begin{axis}[
			ylabel near ticks,
			xlabel near ticks,
			minor xtick={1940, 1942, ..., 1994},
			xtick={1941, 1943, ..., 1993},
			xmin=1939, xmax=1994,
			ymin=-25000, 
			xlabel={Document Years},
			ylabel={Document Count},
			grid=major,
			x tick label style={/pgf/number format/1000 sep=},
			width=.9\textwidth,
			xticklabel style={rotate=90},
			axis lines=left,
			legend style={at={(.62,1.12)},
				anchor=north,legend columns=1,legend cell align={left},draw=none},
			]
			\addlegendentry{\scriptsize Open Access Yearly Document Sum}
			\addplot[thick, color=olive]
			table[x=Godina, y=Broj radova, col sep=comma]{3_OA_Radovi_po_tipu.nlo};\label{1dokOAsupp3}
			\addlegendentry{\scriptsize Subscription Access Yearly Document Sum}
			\addplot[thick, color=teal]
			table[x=Godina, y=Broj radova, col sep=comma]{3_SB_Radovi_po_tipu.nlo};\label{1dokSBsupp3}
		\end{axis}
	\end{tikzpicture}
	\caption{Open Access (OA) and Subscription Access (SA) Documents by Publication Year---Part Three, from 1940 to 1993. Data was acquired from Scopus, which is the "world's largest abstract and citation database of peer-reviewed research literature" \cite{Elsevier2023a}. No filter was set on document type, and as such, this is a complete and up-to-date outlook.}\label{fig:1OaSbSupp3}
\end{figure}

\begin{figure}[t]
	\centering
	\begin{tikzpicture}[]
		\begin{axis}[
			ylabel near ticks,
			xlabel near ticks,
			minor xtick={1990, 1992, ..., 2025},
			xtick={1991, 1993, ..., 2024},
			xmin=1989, xmax=2025,
			ymin=-40000, 
			xlabel={Document Years},
			ylabel={Document Count},
			grid=major,
			x tick label style={/pgf/number format/1000 sep=},
			width=.9\textwidth,
			xticklabel style={rotate=90},
			axis lines=left,
			legend style={at={(.62,1.12)},
				anchor=north,legend columns=1,legend cell align={left},draw=none},
			]
			\addlegendentry{\scriptsize Open Access Yearly Document Sum}
			\addplot[thick, color=olive]
			table[x=Godina, y=Broj radova, col sep=comma]{4_OA_Radovi_po_tipu.nlo};\label{1dokOAsupp4}
			\addlegendentry{\scriptsize Subscription Access Yearly Document Sum}
			\addplot[thick, color=teal]
			table[x=Godina, y=Broj radova, col sep=comma]{4_SB_Radovi_po_tipu.nlo};\label{1dokSBsupp4}
		\end{axis}
	\end{tikzpicture}
	\caption{Open Access (OA) and Subscription Access (SA) Documents by Publication Year---Part Four, from 1990 to 2024. Data was acquired from Scopus, which is the "world's largest abstract and citation database of peer-reviewed research literature" \cite{Elsevier2023a}. No filter was set on document type, and as such, this is a complete and up-to-date outlook. Documents for the year 2024 are the ones with a planned date.}\label{fig:1OaSbSupp4}
\end{figure}
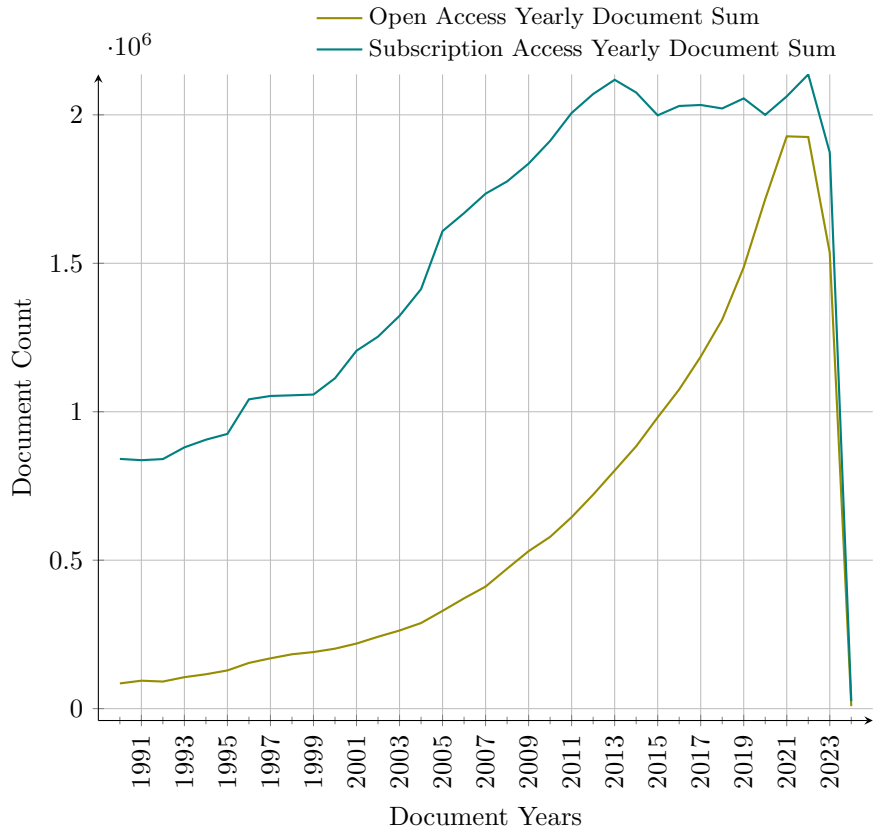

Here we will present a historical overview of the data for Open Access and Subscription Access documents by publication year. Such an image will, in the current significant time, present a useful dataset for future research endeavors and a more insightful outlook on disparate models of scientific publishing for the current research. A visual representation of the data\footnote{The reader can either acquire the raw data for this research from the Scopus database by himself, or he can contact the corresponding author of the article and receive the raw data that way.} in question is presented in Figure~\ref{fig:1OaSbSupp1}, Figure~\ref{fig:1OaSbSupp2}, Figure~\ref{fig:1OaSbSupp3}, and Figure~\ref{fig:1OaSbSupp4}.

The first analysis, presented in Figure~\ref{fig:1OaSbSupp1}, starts from 1865. Even though Scopus covers records from all the way back to 1788 \cite{Elsevier2023b}, the data from which one can make a comparative analysis of Open Access and Subscription Access starts from 1865; therefore, this is the starting year of the analysis. Open Access is not as young\footnote{It should be noted that the Scopus database is a live database and that publishers can update the data. Which means that the further back in time an Open Access entry is, the less likely it is that such a document was Open Access at the time of publication, as a document might have become Open Access in the future and a publisher changed a Scopus data entry.} as one might perceive; however, during years past, it was not as prevalent, and Subscription Access had a substantial lead. Both models of publishing had approximately the same rate of growth. Then something fascinating happened a year or two before World War I: in 1914, there was a substantial decrease in the number of published documents by both models. Open Access seems to have started this decrease earlier, but Subscription Access followed. Then again, in 1918, when World War I ended, a number of documents published again started to have an upward trend; however, in 1922, Open Access suddenly drooped immensely\footnote{One could perhaps argue that such a decline in Open Access was caused by copyright laws. However, copyright laws are far older than the years here in question, starting in England in 1662 with the An Act for preventing the frequent Abuses in printing seditious treasonable and unlicensed Books and Pamphlets and for regulating of Printing and Printing Presses \cite{Nipps2014}, and it does not seem that such laws at the time would produce a change of trend in regard to the state a few years before 1914.}, while Subscription Access went into renaissance. Is it possible that the aftermath of World War I and perturbations on the world stage led to such a decline in Open Access publishing---a pattern repeated throughout history?

In any case, as seen from Figure~\ref{fig:1OaSbSupp2}, after an initial explosion for SA, the growth was lower but steady, while the OA publishing stayed alive, but approximately on the same level and far below SA. Then again, in 1939, the year World War II began, the number of documents published started to decline for both models, with SA this time taking the brunt of the decline. This state of events lasted until 1944, 1945, aligning with the end of World War II in 1945, thus marking another era in scientific publishing. From then onward, SA rose to unseen heights, while OA started its slow but steady incline, as one can observe from Figure~\ref{fig:1OaSbSupp3}\footnote{A supersized increase in scientific research output that required a transformation of then-subsidized and deficit-stricken learned publishers and a strong presence of commercial publishers that were able to transform scientific publishing from a poor investment to a lucrative business. \cite{Fyfe2021}}. By the beginning of the 1990s, the difference between these two models was so vast, and the SA was so dominant, that any competing comparison was not of a serious nature.

This was not the end, however, and if one looks at Figure~\ref{fig:1OaSbSupp4} OA Publishing was in for a serious catching up.\footnote{The official beginning of OA that was done at scale, enabled by digital technology. Prior to that time, OA was, as it were, underground, with sharing done, for example, through email lists.} The OA explosion that started around 1992, 1993, was carried on the wings of the immense expansion of digital technology and the Internet, with 1991 being the year when the well-known arXiv curated by Cornell University was brought into existence \cite{Ginsparg2011}. During the last stretch of time, Open Access has had an ever-increasing growth until the year 2021, with 2022 being of the same nature, and almost equals SA when this growth is suddenly stopped. SA, on the other hand, shows a stagnation from the year 2013, with subsequent years fluctuating. For a long time now, the promotion of OA publishing has been strong, yet as time went by and almost three decades of strong OA growth passed, it became evident that such a model has its own serious flaws and that a change in course is needed. It seems that we came to such a point of change \cite{Sanderson2023}.

The future will perhaps not be Open Access only, but some sort of combination of models. Subscription access and Open Access models are it seems both here to stay, but will this status quo remain, or are we to see another contender or even more contenders? It seems that the future of science will be, in general, more open than traditional SA would produce and less open than OA would produce, while the price will most likely not go substantially down, although the one who pays that price might change.

Figures~\ref{fig:1OaSbSupp1}, \ref{fig:1OaSbSupp2}, \ref{fig:1OaSbSupp3} and \ref{fig:1OaSbSupp4} represent four periods of time of scientific publishing: ca. 55, 30, 45, and 32 years, respectively, as per each figure period. It seems that every 30 to 50 years there is a change in scientific publishing, and it also seems that this gap is shortening. We are currently living in a time when one would expect another change. As for the number of documents published, the first two periods were in the tens of thousands, growing, even substantially, but in the same order of magnitude. Since the end of World War II, publications have skyrocketed into hundreds of thousands, and the world has seen incredible technological progress. The last period speaks in millions, and in a somewhat erratic manner at that. The document count being published is staggering, with many questions that one could inquire, e.g., is the global war-like state of the world of today responsible for the flat peak of OA and erratic behavior of SA observed in Figure~\ref{fig:1OaSbSupp4}? This saturation of documents, however, might also be an indication of things to come. When one remembers unrest in Ukraine in the year 2014 and the events that led to the unrest \cite{OConnell2017,Umland2020}, with the current Israel-Hamas war that began in 2023 \cite{Wynia2023,Hitman2023}, by having in mind historical happenings from 1865 until now, data on the number of documents in Figure~\ref{fig:1OaSbSupp4} is exactly on point. It would therefore be a valid question of research on a dataset that would allow for such a correlation analysis: whether events of significant turmoil, including those rocking the world now, can be statistically linked to trends in scientific publishing.

If OA does not overshoot SA, it is quite possible that regardless of the changes, this will be the model that will stay dominant, at least for the foreseeable future. As models come and go, SA just might be that fixed point that, in combination with other models, stays on top. And so, that which was meant to be for a transitioning period, from SA to OA, might just be that what will not be so easily moved aside and dethroned. Whatever the current situation might be---just a glitch on the horizon followed by OA dominance, or turn of events and SA comeback, or a ground for some other development---it seems that we will soon find out.

It is also worth mentioning that the historical data might be, at least in part, linked to the industrial revolutions, especially the second \cite{Lamoreaux2007} and third \cite{Fitzsimmons1994}, with the fourth \cite{Petrillo2018}, which is sometimes disputed and seen as the continuation of the information age, being more difficult to strongly delineate. Thus, data from Figure~\ref{fig:1OaSbSupp4} might indicate another industrial revolution in the coming years (or are we already at or around the beginning of one?), perhaps in terms of advancements in artificial intelligence, robotics, and quantum computing.

\section{Results}\label{sec4}

Intentions might have been good, and it would not be the first time that good intentions have paved the way to hell, even Open Access hell. As hard as it may seem, the problem is not solved; it has just been transferred from one side to the other \cite{Packer2023}, and there are still those that can't pay, with additional problems added to the fire, namely the explosion of predatory journals like probably never in the history of science, being unable to publish, journals transforming into money-making machines, buying your way into science, etc.

The original problem could have been solved with a surgical knife and with far less time, resources, and effort simply by modifying the subscription model. Was it not easier to waive the subscription or modify the price than to go into the decades-long transformation of scientific publishing \cite{Laakso2011} without resolving the core issue and adding a few new issues into the oven? The scientific ship has, however, sailed, and it is unlikely that it will change course with the decisions reversed, but there is a solution. There are actually at least three solutions\footnote{There is also a possibility of a community-based publishing process that would be entirely scientist- or expert-driven and very inexpensive, yet would that be accepted by the scientific community at large? And what about tenure evaluators, and how well would such a system perform, etc.? \cite{Hutson2021}}, which are easy to implement as they are grounded in everything the publishers already know, with a number of these issues being tackled in the Coalition S analysis also \cite{Packer2023}.

The beginning of wisdom in Full Gold Open Access Publishing is a moment when one realizes that such a model is not, in all its facets, truly open, as the state of the art review and scientific discussion thus far have shown. In the end, what will the one read if the one is unable to publish? The first solution to the issue is to not completely transform into the Full Gold Open Access\footnote{There is also a possibility of having separately Full Gold Open Access journals and Subscription journals, and holding the balance and avenues for scientific publications open in that way.}, or a model alike to it, but to hold two parallel models in place, an Open Access model, and a subscription\footnote{With inability to access specific articles solved by for example posting preprints on places such as arXiv.org curated by Cornell University, Research Square, etc., or by shortening embargo period and making themselves research documents available online, which some journals have done.} model, thus via hybridization holding the flaws of both models in check, a competitive solution that will allow science to flourish. This solution would allow publishers to lower their subscription fees on the account of Open Access articles (which at least some publishers, as far as we are aware, have not yet done \cite{Du2022}).

The second path is lowering the Article Processing Charge to a level, or at least near to, the publication fees employed by scientific conferences, ranging from around $ 200 $ to $ 800 $ dollars. This model would coexist with the standard subscription model (thus potentially having a smaller subscription fee) and would be attractive for the authors, the potential downside being that even though the Article Processing Charge is being paid for a number of papers, no article is immediately free for the reader (a potential consequence; however, it doesn't have to be the case), but with the embargo period for an article being shorter than for a subscription-based article.

The third option is typically called Diamond Open Access \cite{Fuchs2013}. In this instance, the reader can access articles at no charge, while the author publishes the article at no charge, with the journal and ultimately the publisher financing the publishing business from other sources, e.g., commercial endeavors, donations, grants, etc. \cite{Fuchs2013} The challenge here is: why would someone fund this kind of publishing business? This model seems suitable for libraries, universities, the scientific community, or as a certain part of a diversified business model. It is also conceivable that there would be instances where a private company would have an interest in funding a knowledge funnel, so to speak, perhaps to foster its own innovation.

In order to gain an incentive to implement the aforementioned measures in parallel, publishing businesses need to rethink. The problem is not that we have publishing models that are not good, but that we are inhibiting the coexistence of multiple publishing models, thus exacerbating current issues and, in the process, creating new ones. The force by which transition to Full Gold Open Access is being enabled is strong, and if followed by the exclusivity of said model, we will likely find ourselves in an undesirable situation. It is not about one publishing model or the other, but about their coexistence, competition, and their own demise or rise as per the needs of science---competition and demand will prune and remove as necessary. In a world of academy publishing where the pervasive thought is that some exclusive publishing model will solve the issues of science, the thought of suggesting the coexistence of multiple models is revolutionary.

This is bringing us to the last question, and that is, how to make the aforementioned models and measures a reality. There are three avenues that could make it so. The first is the behavior of the primarily scientific community in choosing where to publish and in creating journals and venues where such models and measures will be in place. The second is navigating to a goal through grants for publishing in journals that have in their works these or similar models; grants could also potentially be applied directly to journals that would implement such measures and then apply for a grant. The third is publishers themselves. Publishing is a business, as it should be, and that business will follow the money, but it will also follow another irreplaceable component: editors, reviewers, and academic staff. As difficult as it may be to accept, we are a part of the problem, but the good news is that we are also a part of the solution. It is time to act within journals themselves, and the change will come.

All three designed and suggested models are competitive and balanced, both within themselves and compared to each other, and they will likely best work in parallel, through multiple journals. Such a state of affairs will, as a consequence, make freedom and the flourishing of science a reality until, if necessary, other new or novel models are implemented (while preprint servers can resolve article inaccessibility), with opportunity and a chance for everybody. This will indeed then be the open science and the general model that is sustainable while working well in the long run.

\section{Conclusions}\label{sec5}

The certification service that is being paid for either needs to cost less, or there will be some tumultuous events in academia, or perhaps nothing that would stop this negative trend would happen, and we will indeed end up in a world of pay-or-perish.

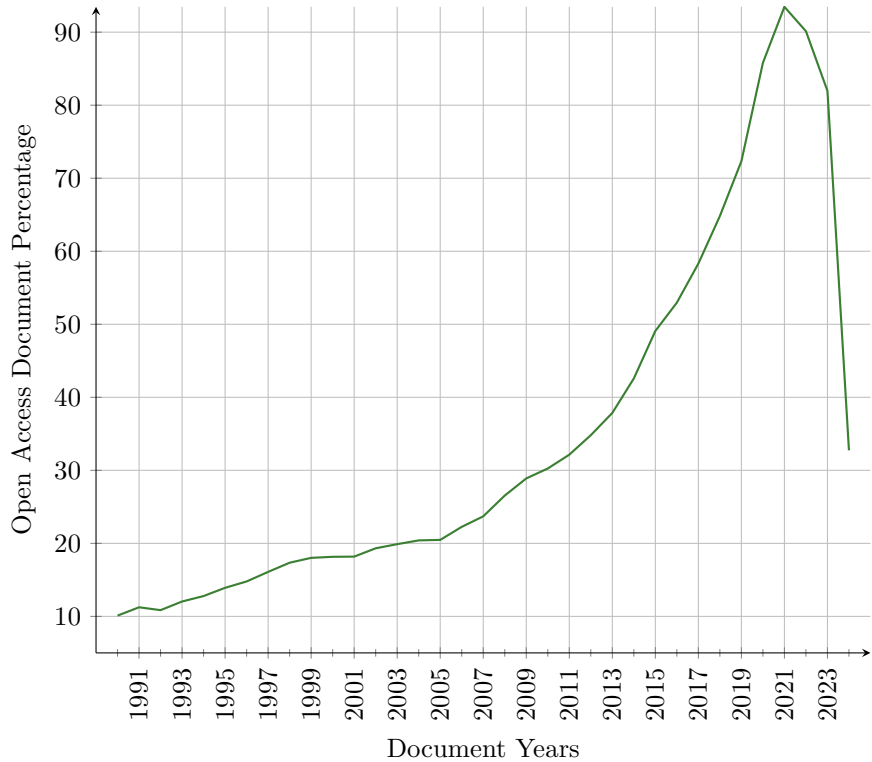
\begin{figure}[h]%
	\centering
	\begin{tikzpicture}[]
		\begin{axis}[
			ylabel near ticks,
			xlabel near ticks,
			minor xtick={1990, 1992, ..., 2024},
			xtick={1991, 1993, ..., 2023},
			xmin=1989, xmax=2025,
			ymin=5, 
			xlabel={Document Years},
			ylabel={Open Access Document Percentage},
			grid=major,
			x tick label style={/pgf/number format/1000 sep=},
			width=.9\textwidth,
			xticklabel style={rotate=90},
			axis lines=left,
			]
			\addplot[thick, color=OliveGreen]
			table[x=Godina, y=Postotak, col sep=comma]{OA_SB_postotak.nlo};\label{leg:pstOA}
		\end{axis}
	\end{tikzpicture}
	\caption{Open Access (OA) as a Percentage of Subscription Access (SA) Documents by Publication Year: The last two years obviously do not have a complete dataset. Data was acquired from Scopus, which is the "world's largest abstract and citation database of peer-reviewed research literature" \cite{Elsevier2023a}. No filter was set on document type, and as such, this is a complete and up-to-date outlook. This analysis accompanies Figure~\ref{fig:OaSb} and presents a relative perspective for the OA model. From a rather small percentage of ca. $ 10\% $, through the years and an ever-increasing popularity, into the last few years and an almost even footing with ca. $ 93\% $ for the year 2021. It seems, however, that some sort of deadlock has been reached, from which the future might proceed in a similar fashion, or there will perhaps be a significant change on the horizon. With various possibilities and models at our disposal, how the future will look is up to us. Documents for the year 2024 are the ones with a planned date.}\label{fig:pstOA}
\end{figure}

Out of the mentioned models, the first and third are likely the most appealing, with this needing to be negotiated in some way; however, as history has shown, publishers are not exactly easy negotiators, as Donald Knuth himself found out when arguing price with the largest scientific publisher. \cite{Knuth2003,Elsevier2010} Which presents one more reason for a multifaceted approach to publishing. Following the same vein, publishers will agree to a model that will ensure thriving, while suggested models of publishing from which one can choose will ensure that thriving will be the case.

If we observe the relation between OA and SA in the Historical Outlook section of the article, it can be seen that the SA model has served science for a long time, and it seems well. OA has a number of advantages; however, its renaissance that started in 1991 has been suddenly interrupted lately, and it just might be that OA will not command the future. It also seems that competition and changing publishing models are not a cause of disturbances in scientific publishing; however, tumultuous world events, it seems, are; at least that is an indication. Such an indication presents itself as a logical consequence of secrecy, unpublished research, newly developed technology, and hidden scientific breakthroughs.

Then, when the disturbance is over, a substantial output in the number of documents published is expected. This has happened during both world wars; it has potentially happened during other periods as well, but perhaps to a lesser extent and more difficult to detect; and it seems to be happening in today's world, in the here and now. The suggested models of publishing and related measures are therefore a sound solution for such a state in science and also for the future. Firstly, because it services science in all its facets, and secondly, because they represent a stable standing ground in a time of uncertainty. Such a stable and historically proven fruitful ground for academic publishing should also be a welcoming ground for innovation, not only strictly classically scientific but also in terms of potential new publishing models.

Consequently, if history is an indication, when the problems settle, a new wave of published documents will ensue. With the valid question of when document saturation happens and are we also in it now, could that be a part of the answer to the current plateau in the number of documents and seemingly stagnation? Things in nature often work in cycles, and it seems that academy publishing is not an exception in this regard. Yet, the future will need to unravel itself before we can claim for sure what the standing of the points the current data reveals is.

The time is fast approaching, with some publishers\footnote{Countries are following similar or the same trend \cite{Zhang2022}.} already announcing that by a certain time, it seems around the year 2025/2026, they will be completely, and others perhaps in a large part, $ 100\% $ Gold Open Access \cite{ACM2023,ACM2023a}---a consequence of which will then be, for those Full Gold Open Access journals, exclusivity of science, either pay the charge, or you will not get published, quite the opposite of the original intent of the openness of science.

As the data in Figure~\ref{fig:OaSb} and Figure~\ref{fig:pstOA} clearly indicate, backed up by the state of the art literature review performed thus far, we are on the verge of consequential events\footnote{Another plight in scientific publishing are sham articles \cite{VanNoorden2023}; it seems that scientific publishing is being "attacked" on multiple fronts, and if something isn't done quickly, integrity together with confidence in research and results will likely suffer substantially.}, and the final goal needs to be modified with an opportunity seized, meritocracy has to be saved \cite{Abbot2023}.

\backmatter

\section*{Declarations}

The author declares no conflict of interest.

\clearpage
\bibliography{sn-bibliography}

\end{document}